%% file: 3rdGen.tex
\documentclass{article}
\usepackage{amsmath}
\usepackage{graphicx}
\usepackage{amssymb}
\usepackage{subfigure}
\usepackage[margin=1in]{geometry}
\usepackage{tabu}
\usepackage{cite}
\usepackage{caption}
\def\beq {\begin{equation}}
\def\eeq {\end{equation}}
\def\bea {\begin{eqnarray}}
\def\eea {\end{eqnarray}}
\def \PMET{\rm p{\!\!\!/}_T}
\def \MET{\rm E{\!\!\!/}_T}

\def \PMETV{\overrightarrow{\rm p}{\!\!\!\!/}_T}
\newcommand{\br}{\begin{eqnarray}}
\newcommand{\er}{\end{eqnarray}}
\newcommand{\be}{\begin{equation}}
\newcommand{\ee}{\end{equation}}

\usepackage{color}
\usepackage{hyperref} 
\hypersetup{linktocpage=true}
\usepackage[all]{hypcap}
\hypersetup{
   bookmarks=true,         
   unicode=false,          
   pdftoolbar=true,        
   pdfmenubar=true,        
   pdffitwindow=false,     
   pdfstartview={FitH},    
   pdftitle={My title},    
   pdfauthor={Author},     
   pdfsubject={Subject},   
   pdfcreator={Creator},   
   pdfproducer={Producer}, 
   pdfkeywords={keyword1} {key2} {key3}, 
   pdfnewwindow=true,      
   colorlinks=true,       
   linkcolor=blue,          
   citecolor=magenta,        
   filecolor=magenta,      
   urlcolor=cyan           
}
\usepackage{fancyhdr}
\begin{document}
\thispagestyle{empty}
\vspace*{-22mm}
\vspace*{10mm}
\hypersetup{backref=true,bookmarks}

\vspace*{10mm}

\begin{center}
{\Large {\bf\boldmath 
Stop and sbottom search using dileptonic $\rm M_{T2}$ variable 
and boosted top technique at the LHC}}

\vspace*{10mm}

{\bf Amit Chakraborty$\rm ^a$, Dilip Kumar Ghosh$\rm ^a$, 
Diptimoy Ghosh$\rm ^b$, Dipan Sengupta $\rm ^c$}\\
\vspace{4mm}
{\small

$\rm ^a$ Department of Theoretical Physics, Indian Association for the 
Cultivation of Science, \\ 
2A \& 2B, Raja S.C.\,Mullick Road, Jadavpur, Kolkata 700\,032, India\\[2mm]

$\rm ^b$ INFN, Sezione di Roma,\\
Piazzale A.  Moro 2, I-00185 Roma, Italy \\ [2mm]

$\rm ^c$ Department of High Energy Physics, Tata Institute of 
Fundamental Research,\\
1, Homi Bhabha Road, Mumbai-400\,005, India}

\vspace*{10mm}

{\bf Abstract}\vspace*{-1.5mm}\\
\end{center}

The ATLAS and CMS experiments
 at the CERN LHC have collected about 25 $\rm fb^{-1}$ of data each
 at the end of their 8 TeV run, and ruled out a huge swath of parameter
 space in the context of Minimally Supersymmetric Standard Model (MSSM).
 Limits on masses of the gluino ($\rm \tilde{g}$) and the squarks of
 the first two generations ($\rm \tilde{q}$) have been pushed to above
 1 TeV. Light third generation squarks namely stop and sbottom of
 sub-TeV masses, on the other hand, are still allowed by their direct
 search limits. Interestingly, the discovery of a Standard Model (SM)
 higgs boson like particle with 
a mass of $\sim$ 125 GeV favours a light third generation which is also 
motivated by naturalness arguments. Decays of stop and sbottom quarks 
can in general produce a number of distinct final states which 
necessitate different search strategies in the collider experiments. 
In this paper we, on the 
other hand, propose a general search strategy to look for third 
generation squarks in the final state which contains a top quark in the 
sample along with two additional hard leptons and substantial missing 
transverse momentum.  We illustrate that a search strategy using the 
dileptonic $\rm M_{T2}$, the effective mass $\rm m_{eff}$ and jet 
substructure to reconstruct the hadronic top quark can be very effective 
to reduce the SM backgrounds.  With the proposed search strategy, we 
estimate that the third generation squarks with masses up to about 900 
GeV can be probed at the 14 TeV LHC with 100 $\rm fb^{-1}$ luminosity. 
We also interpret our results in two simplified scenarios where we 
consider the stop (sbottom) pair production followed by their subsequent 
decay to a top quark and the second lightest neutralino (lightest 
chargino). In this case also we find that stop (sbottom) mass up to 1 
TeV (0.9 TeV) can be discovered at the 14 TeV LHC with 100 $\rm fb^{-1}$ 
integrated luminosity.

\vspace*{10mm}


\noindent

\newpage

\tableofcontents

\section{Introduction and motivation}
\label{intro}

As the LHC suspends its operation for the upgrade to 14 TeV after having 
collected about 25 $\rm fb^{-1}$ of data by each of the
 ATLAS\cite{Aad:2008zzm} and CMS\cite{Chatrchyan:2008aa}  
experiments at the centre of mass energy of 8 TeV, we look back and 
marvel at the amazing performance of the machine and the breakneck pace 
at which the ATLAS and CMS collaborations have been analyzing the data. 
The feather in the cap of the 8 TeV run has been the most anticipated 
result in particle physics in the last two decades; the discovery of a 
boson of mass about 125 GeV which resembles the standard model(SM) higgs 
boson in its behavior \cite{:2012gu,:2012gk}.

The discovery of a 125 GeV higgs like boson has fueled speculations of 
physics beyond the SM from various considerations like vacuum stability, 
fine tuning of the higgs potential and so on. The impact of this 
discovery on new physics scenarios has been studied extensively in the 
recent past. Its implications on Supersymmetry(SUSY) have been 
discussed in a wide range of papers 
\cite{Baer:2011ab,Akula:2011aa,Feng:2011aa,Heinemeyer:2011aa, 
Buchmueller:2011ab,Draper:2011aa,Cao:2011sn,Hall:2011aa,Ellis:2012aa, 
Baer:2012uya,Maiani:2012ij,Cheng:2012np,Cao:2012fz,Brummer:2012ns, 
Balazs:2012qc,Feng:2012jf,Ghosh:2012dh,Fowlie:2012im,Athron:2012sq, 
CahillRowley:2012rv,Akula:2012kk,Cao:2012yn,Arbey:2012dq,Nath:2012fa, 
Ellis:2012nv,Chakraborti:2012up,Chakraborty:2013si, 
Mayes:2013qmc,Dighe:2013wfa}. Along with this there have been sustained 
efforts in the particle physics community to probe SUSY signatures in 
the light of the recent higgs data 
\cite{Ghosh:2012mc,Choudhury:2012tc,Byakti:2012qk, 
Chatterjee:2012qt,Baer:2012vr,Howe:2012xe,Ghosh:2012wb,Ghosh:2012ep,Arbey:2012fa,
Bhattacherjee:2012bu,Ghosh:2013qga,Belanger:2013oka}.
While after the first two years of data taking the gluinos and first two 
generation of squarks do not seem to be in the sub-TeV scale, the discovery 
of the higgs like boson has spurred a renewed interest in a region of 
SUSY parameter space termed as ``natural Supersymmetry'' 
\cite{Hall:2011aa,Feng:2012jf,Berger:2012ec,Cao:2012rz,Randall:2012dm, 
Espinosa:2012in}. This idea stems from the fact that the most relevant 
superparticles responsible for cancellation of the quadratic divergence 
in the higgs mass are the third generation squarks. Therefore, even if 
squarks of the first two generations and the gluinos are heavy, a 
comparatively lighter third generation can cure the fine-tuning problem 
in the SM. While a 125 GeV higgs would, in general, necessarily mean a 
large stop tri-linear coupling $\rm A_t$, introducing some adjustment of 
parameters in the theory, a light third generation scenario nevertheless 
has an extremely attractive prospect for both the theorists and the 
experimentalists. This has brought about a paradigm shift in the way 
SUSY searches have been conducted at the LHC.

At the LHC the SUSY searches have been performed quite extensively both in the framework 
of constrained minimally supersymmetric standard model (cMSSM) which 
assumes specific relations among the soft supersymmetry breaking 
couplings at a very high energy scale and in other models which relax these 
simplifying assumptions to various degrees. There have been quite a few 
phenomenological analysis to investigate cMSSM signatures at the 
LHC using 
different techniques to suppress SM 
background\cite{Ghosh:2012dh,Baer:2012vr,Chatterjee:2012qt}. The limit 
on the gluino mass obtained by ATLAS and CMS collaborations in the cMSSM 
context currently stands at about 1.5 TeV for $\rm m_{\widetilde{g}}\simeq 
m_{\widetilde{q}}$ and about 1.2 TeV for $\rm m_{\widetilde{g}} \ll 
m_{\widetilde{q}}$\cite{CMS-PAS-SUS-12-005,ATLAS-CONF-2012-109}.
More recently, however, other models specialy R-parity 
violating SUSY models and gauge mediated supersymmetry breaking scenarios
have also been studied by both ATLAS\cite{ATLAS-CONF-2013-026,ATLAS-CONF-2013-036} and 
CMS\cite{CMS-PAS-SUS-13-010,CMS-PAS-SUS-13-003} collaborations.

While cMSSM searches have provided a guideline to the way collider 
strategies are devised at the LHC, the negative results in all the 
searches have triggered a more pragmatic approach to SUSY searches at 
the LHC. This together with the discovery of a SM higgs like boson has 
lead us to look beyond cMSSM, in particular at considering simplified 
models motivated by naturalness arguments and hence, light third 
generation sparticles in a general MSSM framework with no specific 
relations among the soft supersymmetry breaking parameters. This has 
motivated an avalanche of phenomenological studies in light stops and/or 
light sbottom scenarios 
\cite{Desai:2011th,He:2011tp,Ghosh:2012wb,Drees:2012dd,Berger:2012ec,Plehn:2012pr, 
Han:2012fw,Barger:2012hr,Choudhury:2012kn,Cao:2012rz,Chen:2012uw, 
Bornhauser:2010mw,Kraml:2005kb,Yu:2012kj,Ajaib:2011hs,Ghosh:2012ud, 
Dutta:2012kx,Alves:2012ft,Berenstein:2012fc,Kilic:2012kw}. In terms of collider 
searches most phenomenological studies have been conducted for 
predominantly right handed stop and sbottom quarks in the decay channels 
$\rm \widetilde{t}\to t \widetilde{\chi}_{1}^{0}$ and $\rm 
\widetilde{b}_{1}\to b \widetilde{\chi}_{1}^{0}$ 
\cite{AdeelAjaib:2011ec,Ajaib:2011hs,Lee:2012sy,Alvarez:2012wf,Ajaib:2012eb}. 
The general observation has been that the efficiency depends strongly on 
the $\rm \widetilde{t}_{1}(\widetilde{b}_{1})- \widetilde{\chi}_{1}^{0}$ 
mass difference which, if large, leads to a hard jets and a large 
missing transverse energy ($\rm \MET$) crucial in suppressing the 
large background arising from the production of top quarks in the SM, 
while for smaller mass difference scenarios it has been 
suggested to look for stop pair production in association with one hard QCD jet 
prompting monojet + $\MET$ final state\cite{He:2011tp,Drees:2012dd}. 
Some analyses have 
tried to overcome this problem by performing a shape analysis of the 
distributions of $\MET$ and $\MET$-related observables 
\cite{Alves:2012ft}. It has also been suggested that in the large $\rm 
\widetilde{t}_{1}-\widetilde{\chi}_{1}^{0}$ mass difference scenario, 
the use of top-tagging by jet substructure methods can also be very 
helpful in achieving a reasonable sensitivity both from direct stop 
production and in stops from gluino cascade decays 
\cite{Plehn:2010st,Kaplan:2012gd,Berger:2011af,Ghosh:2012ud,Bandyopadhyay:2010ms}.

Recently several phenomenological studies have been performed 
to investigate the signature of strongly interacting third generation 
SUSY particles in various channels at the 14 TeV run of the LHC. 
A monojet with large $\MET$ signal has been proposed to 
look for the light stop, where $\rm \widetilde t_1 \to c \widetilde\chi^0_1 $ 
via one loop processes. Using this signal topology, it has been found 
with 100 ${\rm fb}^{-1}$ luminosity a light stop mass up to 250 GeV 
can be probed \cite {Drees:2012dd}.
Similarly the prospect of a spin zero top partner decaying to top quark and 
neutral particles was also studied in \cite{Chen:2012uw} and the 
expected discovery limit for light stop mass is close to 675 GeV 
with 100 $\rm fb^{-1}$ luminosity.
The gluino-stop-neutralino model has been 
also studied in \cite{Ghosh:2012ud} with two top tagged 
jets and missing energy. The resultant gluino discovery 
limit is expected to be 1.7 TeV with 33 $\rm fb^{-1}$ luminosity.
As far as the sbottom searches are concerned, it has been shown
that in the sbottom-co-annihilation scenario 
$\rm \widetilde{b}_{1} \to b \tilde{\chi}_{1}^{0}$ 
the light sbottom mass up to 570 GeV can be observed 
once again with 100 ${\rm fb}^{-1}$ data
\cite{AdeelAjaib:2011ec}.

As far as the experimental efforts to look for the signal of
stop and sbottom at the LHC are concerned, both the ATLAS and 
the CMS Collaborations have searched for light 
stops in simplified models and placed limits assuming specific mass 
relations among gluino, stop and the lightest neutralino. The CMS
collaboration analyzed about 10.5 $\rm fb^{-1}$ of data set at 8 TeV 
center of mass energy in 
the same sign di-lepton + b-tagged jets final state and placed limits in 
various simplified model scenarios\cite{:2012vh}. This study ruled out 
up to 1 TeV gluino mass in a model where the gluino decays via an 
off-shell stop, $\rm \widetilde{g} \to  t \widetilde t_1^{*} \to t \bar{t} 
\widetilde{\chi}_{1}^{0}$, for the lightest neutralino mass up to 600 
GeV. 
The same study also excluded gluino mass up to 1 TeV, where gluino decays
via on shell top and stop to the same final state as above for 
$m_{\tilde t_1} \simeq 800$ GeV and $m_{\tilde \chi^0_1} = 50$ GeV.
They also examined models where the gluino cascades via on-shell sbottoms to 
charginos and finally to di-leptons ($\rm \widetilde{g} \to 
\widetilde{b} b \to t \widetilde{\chi}_{1}^{\pm} b $). This study ruled 
out gluino masses in the range 300 GeV -- 1 TeV for sbottom masses in 
the same range assuming $\rm 
\widetilde{\chi}_{1}^{0}$ and $\widetilde{\chi}_{1}^{\pm}$ masses to be 
50 GeV and 150 GeV respectively. Finally, from the direct sbottom production 
process sbottom masses in the window 270 -- 450 GeV for $\rm 
\widetilde{\chi}_{1}^{\pm}$ masses in the range 100 -- 200 GeV with a 
$\rm \widetilde{\chi}_{1}^{0}$ mass of 50 GeV are excluded. 
The ATLAS collaboration recently searched 
for light stops in the channel $\rm \widetilde{t}_{1}\to b 
\widetilde{\chi}_{1}^{\pm} \to b \widetilde{\chi}_{1}^{0}ff'$ assuming 
$\rm \Delta m \equiv 
m_{\widetilde{\chi}_{1}^{\pm}}-m_{\widetilde{\chi}_{1}^{0}}$ to be 5 GeV 
and 20 GeV. For $\rm \Delta m = 5~GeV$ they ruled out stop masses of 
about 600 GeV in a corridor of lightest neutralino mass 
\cite{ATLAS-CONF-2013-001}.
ATLAS also studied light stops in the decay mode $\rm \widetilde{t}_{1}\to \widetilde{b}\chi_{1}^{\pm}$ 
and $\rm \widetilde{t}_{1}\to t\widetilde{\chi}_{1}^{0}$ with 20.7 $\rm fb^{-1}$ of data in the final state 
with a single lepton in association with jets(including one b-tagged jet) and $\PMET$ 
\cite{ATLAS-CONF-2013-037} using the kinematic variable $\rm M_{T2}$. 
Assuming 100\% branching ratio for $\rm \widetilde{t}_1\to t\widetilde{\chi}_{1}^{0}$ this study ruled out 
light stop masses between 200 and 610 GeV assuming massless LSP, while a stop 
of up to 500 GeV 
was ruled out for LSP masses up to 250 GeV. In the $\rm \widetilde{t}_1\to b\widetilde{\chi}_{1}^{\pm}$
channel, this study ruled out stop masses up to 410 GeV assuming a chargino mass of 150 GeV. 
In the no lepton + jets+ $\PMET$ final state ATLAS looked for stops and sbottoms
in the channels $\rm \widetilde{t}_{1}\to b\widetilde{\chi}_{1}^{\pm}$, and $\rm \widetilde{b}_{1}\to b\widetilde{\chi}_{1}^{0}$ 
with the use of the contransverse mass variable $\rm m_{CT}$ \cite{ATLAS-CONF-2013-053}.
This study ruled out sbottom masses up to 620 GeV for a lightest neutralino mass below 150 GeV. 
The same study also ruled out stop masses up to 580 (440) GeV for 
$\rm \Delta m = m_{\widetilde{\chi}_{1}^{\pm}}-m_{\widetilde{\chi}_{1}^{0}}=5
(20)$ GeV assuming a LSP mass of 100 GeV. Along with these there have been 
studies in the di-lepton channel for stops decaying via the above mode\cite{ATLAS-CONF-2013-048},
while stops in other model scenarios like gauge mediated supersymmetry breaking scenario(GMSB) 
has also been 
considered \cite{ATLAS-CONF-2013-025}. Stop searches by ATLAS at 7 TeV LHC in all the above 
mentioned channels 
also provided limits in this respect \cite{Aad:2012yr,Aad:2012tx,Aad:2012uu}.
Similar studies have also been performed by the CMS collaborations and the most updated results 
can be found in Ref.\cite{CMS-PAS-SUS-13-011,CMS-PAS-SUS-13-008}.

\begin{figure}[ht!]
\begin{tabular}{c}
\fbox{\includegraphics[scale=0.38]{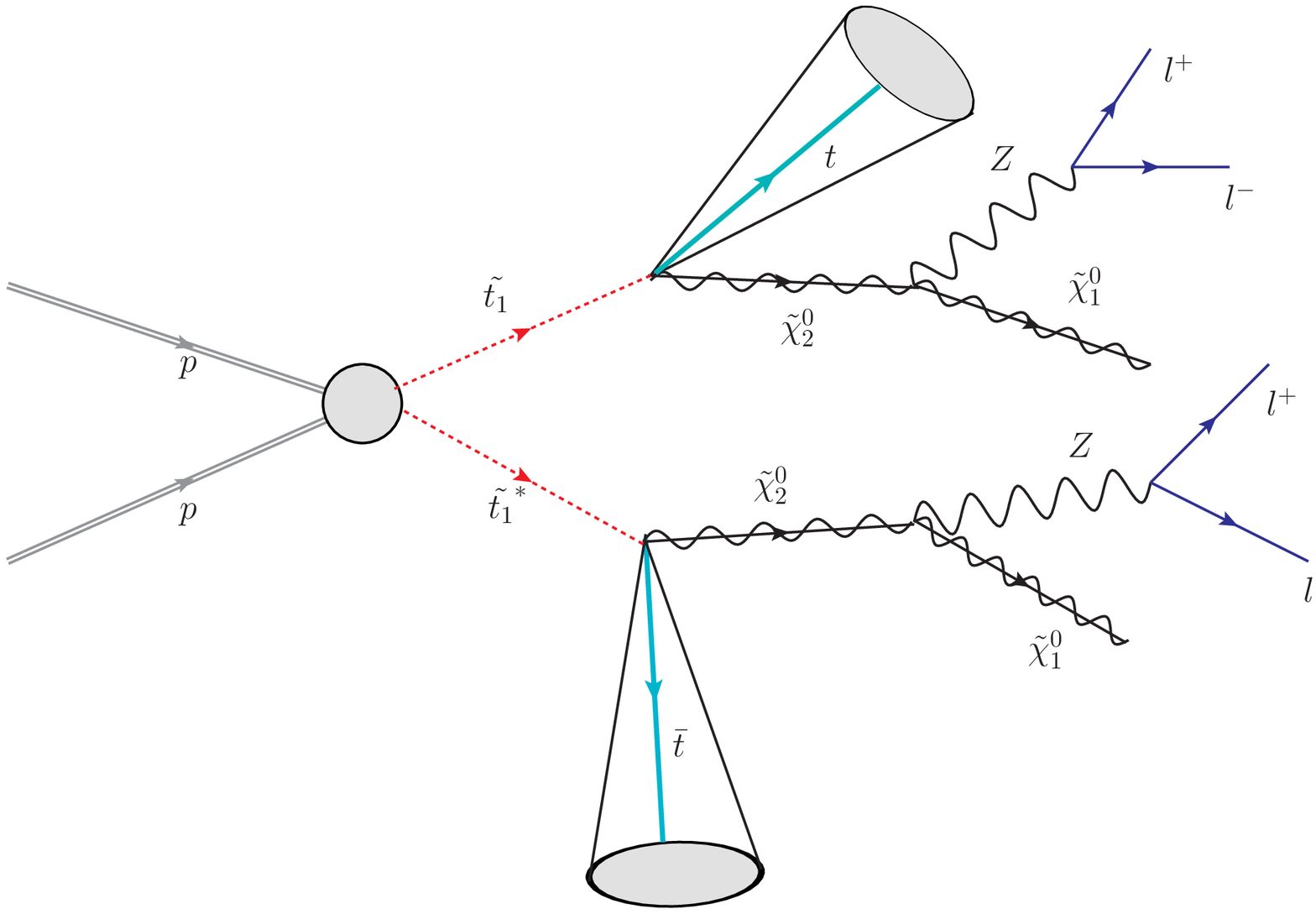}}
\fbox{\includegraphics[scale=0.38]{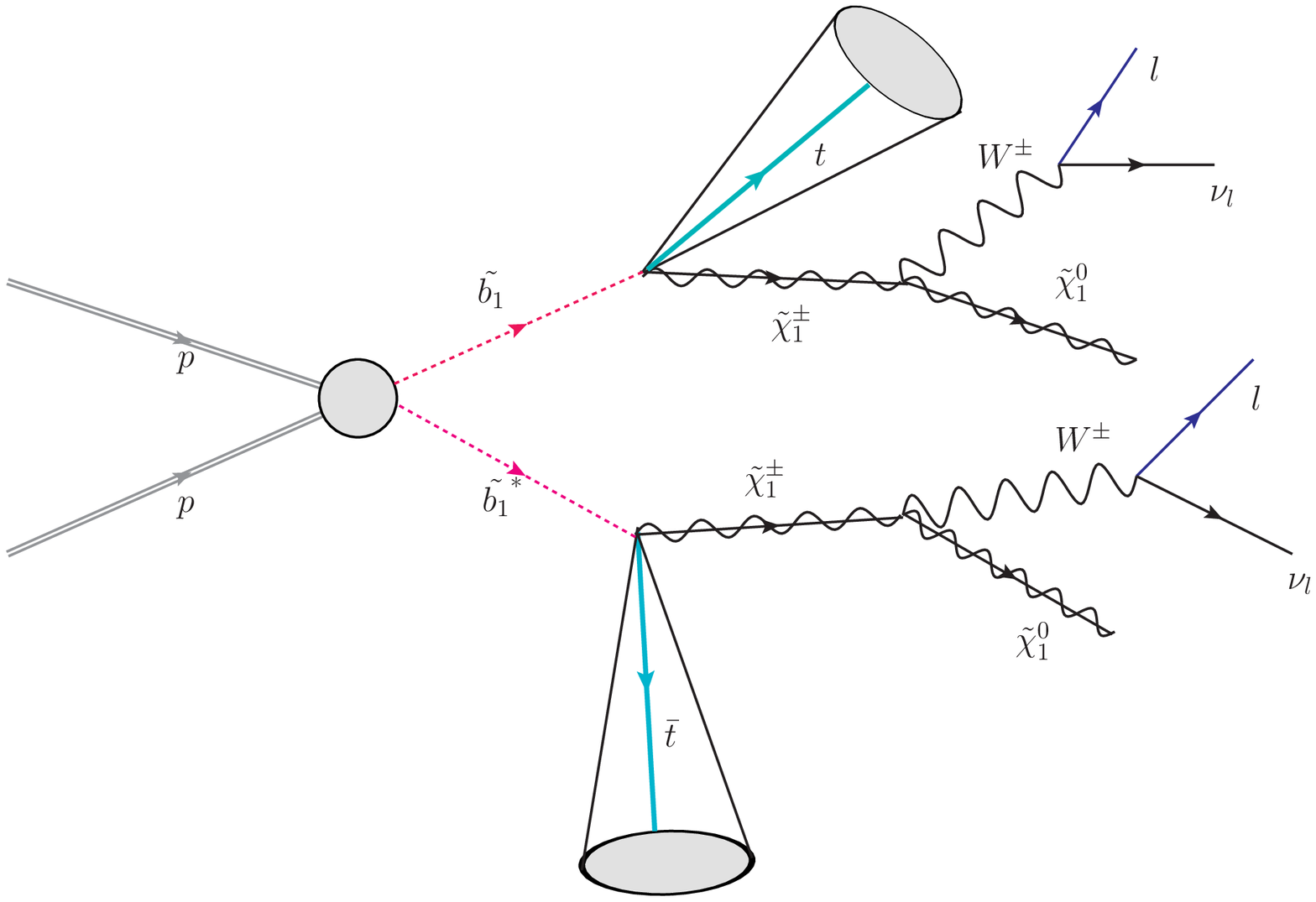}}
\end{tabular}
\caption[]{Sample Feynman diagrams for stop and sbottom pair productions 
and their subsequent decays to a final state containing a top quark, 
two additional leptons and missing transverse momentum. 
\label{fd}}
\end{figure}

It is imperative for some of these search strategies to be extended and 
made more efficient to cover more ground in the quest for 
strongly interacting third 
generation SUSY. Among the various final states that can be analyzed, 
it is also worthwhile to consider a general strategy for stop and 
sbottom quarks such that the same final state is common to both the stop 
and sbottom pair production. Although, light stops and/or light sbottoms 
around or below the TeV scale is well motivated, production cross 
sections for individual stop pair and sbottom pair process 
are rather small even at 14 TeV LHC and thus, the option of an inclusive 
stop and sbottom search is lucrative, as was the case for early 
inclusive cMSSM searches at the LHC for squarks and the gluino. This motivates 
us to consider the final state containing a hadronic top quark, two 
additional hard leptons and missing transverse momentum which can 
originate from the decays of both the stop and sbottom pairs. Some of 
the decay chains which can give rise to such final state are shown 
below,

$$
\begin{array}{ccccccccc}
\rm p  p & \to &  \rm \widetilde{t}_{1} \, \widetilde{t}_{1}^{*} & \to 
&  \rm t \, \bar{t} \, \widetilde{\chi}_{2}^{0} \, 
\widetilde{\chi}_{2}^{0} & \to &  \rm t \, \bar{t}+ 2 Z +
2 \widetilde{\chi}_{1}^{0}   
& \to 
&  \rm t/\bar{t} + \ell \, \ell + \PMET + X  \\
\rm p  p & \to &  \rm \widetilde{t}_{1} \, \widetilde{t}_{1}^{*} & \to 
&  \rm t \, \bar{b} \, \widetilde{\chi}_{2}^{0} \, 
\widetilde{\chi}_{1}^{-} & \to &  \rm t \, \bar{b}+ W^{-} Z +
2 \widetilde{\chi}_{1}^{0}   
& \to 
&  \rm t + \ell \, \ell + \PMET + X  \\
\rm p  p & \to &  \rm \widetilde{t}_{1} \, \widetilde{t}_{1}^{*} & \to 
&  \rm t \, \bar{b} \, \widetilde{\chi}_{2}^{0} \, 
\widetilde{\chi}_{1}^{-} & \to &  \rm t \, \bar{b}+ W^{-} h +
2 \widetilde{\chi}_{1}^{0}   
& \to 
&  \rm t + \ell \, \ell + \PMET + X  \\
\rm p  p & \to & \rm \widetilde{b}_{1} \, \widetilde{b}_{1}^{*} & \to 
&  \rm t \, \bar{t} \, \widetilde{\chi}_{1}^{+}\chi^{-}_{1} & \to 
&  \rm t \, \bar{t} + W^{+} W^{-} + 2 \chi_{1}^{0}  
&\to
&   \rm t/\bar{t} + \ell \, \ell + \PMET + X \\
\rm p  p & \to & \rm \widetilde{b}_{1} \, \widetilde{b}_{1}^{*} & \to 
&  \rm t \, \bar{b} \, \widetilde{\chi}_{1}^{-} \chi^{0}_{2} & \to 
&  \rm t \, \bar{b} + W^{-} Z + 2 \chi_{1}^{0}  
&\to
&   \rm t + \ell \, \ell + \PMET + X \, .
\end{array}
$$
where, $\ell = e~{\rm and}~\mu $.

In Fig.~\ref{fd} we also show two example Feynman diagrams of the 
processes of our interest. Note that, if the stops and sbottoms are 
moderately heavy with a large mass gap $\rm \widetilde{b}_1 - 
\widetilde{\chi}_{1}^{\pm}$ and $\rm \widetilde{t}_{1} - 
\widetilde{\chi}_{2}^{0}$, the produced top quark can be sufficiently 
energetic so that the jet substructure techniques can be very useful to 
reconstruct them in the hadronic channel. Along with this, two 
additional hard leptons with moderately large missing transverse energy 
can lead to a clean signal. In addition to reconstructing a boosted top 
quark, a hard cut on $\rm M_{T2}$ constructed out of the momenta of the 
two hard leptons and the missing transverse energy helps us to combat 
the background in an efficient way.

We organize our paper as follows. In section \ref{model} we discuss the 
choice of our model parameters. A few benchmark points are also selected 
for the illustration of our collider strategy. Section \ref{collider} 
describes the details of our event selection cuts. The signal, 
backgrounds and statistical significance are also shown. The 
interpretation of our results in two simplified models is presented in 
section \ref{simplified} and finally, we summarize our observations and 
conclude in section \ref{concl}.

\section{Choice of model parameters, benchmarks and branching ratios}
\label{model}

In models like cMSSM where all the scalar soft masses are set to a 
common value at a high scale, the lightest stop quark, because of the 
large Renormalization Group(RG) effect, becomes predominantly right handed. 
A bino-like lightest 
neutralino also often emerges from the minimal supergravity boundary 
conditions. Because of this, light third generation searches so far have 
been mostly carried out in the scenario where the dominant decay modes 
for the top and bottom squarks are $\rm \widetilde{t}_{1}\to 
t\chi_{1}^{0}$ and $\rm \widetilde{b}_{1}\to b\widetilde{\chi}_{1}^{0}$ (Provided 
$\rm \widetilde{t}_1-\widetilde{\chi}_{1}^{0}$ mass gap is sufficient for this decay 
to be kinematically allowed).

However, decays of top and bottom squarks to the heavier neutralinos and 
charginos are quite motivated in the context of natural SUSY spectrum 
and have gained considerable attention recently 
\cite{Graesser:2012qy,Dutta:2013sta}. The branching ratios to these 
channels can also be quite high in many 
scenarios\cite{Graesser:2012qy,Dutta:2013sta}. In the context of the 
phenomenological MSSM (pMSSM) this can easily be achieved by having 
the lighter stop and sbottom predominantly left handed. In this case the decays of 
stop and sbottom will be dominated by $\rm \widetilde{t}_{1} \to t 
\widetilde{\chi}_{2}^{0}$, $\rm \widetilde{t}_{1} \to b 
\widetilde{\chi}_{1}^{+}$ and $\rm \widetilde{b}_{1} \to b 
\widetilde{\chi}_{2}^{0}$, $\rm \widetilde{b}_{1}\to t 
\widetilde{\chi}_{1}^{-}$ respectively.  With the increasing masses of 
the top and bottom squarks the modes $\rm \widetilde{t}_{1} \to t 
\widetilde{\chi}_{3,4}^{0}$, $\rm \widetilde{t}_{1} \to b 
\widetilde{\chi}_{2}^{+}$ and $\rm \widetilde{b}_{1} \to b 
\widetilde{\chi}_{3,4}^{0}$, $\rm \widetilde{b}_{1}\to t 
\widetilde{\chi}_{2}^{-}$ also open up. While these latter decay modes 
allow a plethora of final states which can be studied individually, all 
of them also contribute to the final state of our interest. Thus, our 
analysis also captures some of the events arising from these decay 
chains.

In order to perform a detailed analysis of signal and background which 
we present in the next section, we now choose a few model benchmark 
points. Note that both the ATLAS and CMS collaborations have searched 
for electroweak gauginos in the channel 
$\rm p p\to \widetilde{\chi}_{1}^{\pm}\widetilde{\chi}_{2}^{0} \to
W\widetilde{\chi}_{1}^{0} Z\widetilde{\chi}_{1}^{0}$
which have ruled out a range $\widetilde{\chi}_{1}^{\pm}=\widetilde{\chi}_{2}^{0}$ masses as a function of 
the LSP mass \cite{ATLAS-CONF-2012-154,CMS-PAS-SUS-12-022,ATLAS-CONF-2013-035}.
Here, we would like to emphasize that in their analysis the second neutralino 
is always assumed to decay exclusively to the LSP and the Z boson and hence the limits 
obtained there will not apply directly for the case where the second neutralino has 
non zero branching ratio to the higgs boson also. 
In order to be more general, we will consider two different 
scenarios: in one case the dominant decay mode of the second lightest neutralino 
is via the Z boson (benchmarks P2, P3, P6) and in the other case the second lightest 
neutralino decays dominantly via the higgs (benchmarks P1, P4, P5). 
In order to do that, the gluino mass parameter $\rm M_{3}$ is set 
to 1.5 TeV as it is irrelevant for the parameter 
space of our interest and the values of $\rm M_{1}$ and $\rm M_{2}$ are varied,
hence providing various values of $\rm \widetilde{\chi}_{i}^{0} $ and 
$\rm \widetilde{\chi}_{i}^{\pm}$ 
(see Table \ref{tab-1}) to obtain various decay scenarios so as to make our 
search strategy fairly generic. 
The higgsino mass parameter $\mu$ is taken to be 300 GeV and tan$\beta$, 
the ratio of the vacuum expectation values of the two higgs doublets, is fixed at the 
value of 10.

The right handed third generation squark mass parameters $\rm 
m_{\widetilde{t}_{R}}$ and $\rm m_{\widetilde{b}_{R}}$ are set to a high 
value of 2 TeV in order to keep the lightest stop and sbottom mostly 
left handed facilitating the dominance of the decay modes of our 
interest. All the tri-linear couplings, with the exception of $\rm A_t$, 
is set to zero. $\rm A_{t}$ is taken to have a large negative value of 
-2800 GeV just to make sure that the lightest higgs mass is in the range 
123 -- 128 GeV. These tri-linear couplings, however, have very little 
bearing on our final results. The masses of the first two generations of 
squarks and all the three generations of sleptons are set to 5 TeV as 
they are irrelevant for our study. 

To generate the physical masses of the sparticles we have used the 
software package SuSpect \cite{Djouadi:2002ze}. The decay branching 
ratios are then calculated using SUSYHIT\cite{Djouadi:2006bz} which also includes 
SuSpect inside it.
In Table \ref{tab-1} we present the masses of top and bottom squarks along
with the neutralino, chargino states  and the relevant branching ratios for the six benchmark 
points which we choose for our signal and background analysis in the next section.  
The light stop and sbottom masses are obtained by varying the left handed third 
generation squark mass parameter $\rm m_{\widetilde{Q_3}}$ as shown in 
the first column of Table \ref{tab-1}.
As we have already mentioned earlier, in order to demonstrate that our search strategy is 
democratic we choose benchmark points such that $\widetilde{\chi}_{2}^{0}$ decays via both the Z boson 
and/or the Higgs boson. 
Along with these we also provide the cross section for 
the $\widetilde{\chi}_{1}^{\pm}\widetilde{\chi}_{2}^{0}$ pair production at 8 TeV such 
that a rough estimate of the experimental bounds can be checked.

\begin{table}[ht!]
\small
\begin{center}
\tabulinesep=1.2mm
\begin{tabu}{|c|c|c|c|c|c|c|}
\hline
  &  P1 & P2 & P3 & P4 & P5 & P6  \\
\hline
\hline
$\rm m_{\widetilde{Q_3}}$   &  500 & 500 & 700  & 700 & 900 & 900  \\
\hline
 $\rm m_{\widetilde{t}_{1}}$  & 501.7 & 501.7 & 714.2  & 714.2 & 918.1 & 918.1  \\
\hline
$\rm m_{\widetilde{b}_{1}}$  & 525.4 & 525.4 & 748.4 & 748.4 & 918.1 & 918.1  \\
\hline
$\rm m_{\widetilde{\chi}_{1}^{0}}$  & 48.5 & 97.9 & 146.3 & 97.8 & 149.0 & 198.3  \\
\hline
$\rm m_{\widetilde{\chi}_{2}^{0}}$  & 193.3 & 193.9 & 245.9 & 244.3 & 297.9 & 298.6  \\
\hline
$\rm m_{\widetilde{\chi}_{3}^{0}}$  & 309.3 & 309.0 & 309.2 & 309.4 & 408.1 & 408.0  \\
\hline
$\rm m_{\widetilde{\chi}_{4}^{0}}$  & 338.5 & 339.2 & 364.5 & 363.9 & 440.5 & 441.2  \\
\hline
$\rm m_{\widetilde{\chi}_{1}^{\pm}}$  & 192.8 & 192.8 & 242.7 & 242.7 & 297.0 & 297.0  \\
\hline
$\rm m_{\widetilde{\chi}_{2}^{\pm}}$  & 338.9 & 338.9 & 363.5 & 363.5 & 439.8 & 439.8  \\
\hline
$\rm BR(\widetilde{b}_{1} \to b \, \widetilde{\chi}_{2,3,4}^{0}) (\%)$ 
& 34.6 & 34.5 & 19.3 & 19.4 & 19.4 & 19.4  \\
\hline
$\rm BR(\widetilde{b}_{1}  \to t \,\widetilde{\chi}_{1,2}^{\pm}) (\%)$  
& 65.4 & 65.5 & 80.7 & 80.6 & 80.6 & 80.6  \\
\hline
$\rm BR(\widetilde{t}_{1} \to t \, \widetilde{\chi}_{2,3,4}^{0}) (\%)$  
& 34.9 & 35.2 & 62.5 & 62.4 & 62.5 & 62.5  \\
\hline
$\rm BR(\widetilde{t}_{1} \to b \, \widetilde{\chi}_{1,2}^{\pm})(\%)$ 
& 65.1 & 64.8 & 37.5 & 37.6 & 37.5 & 37.5  \\
\hline
$\rm BR(\widetilde{\chi}_{2}^{0} \to \widetilde{\chi}_{1}^{0} \rm{Z})(\%)$ 
& 33.9 & 100.0 & 100.0 & 22.1 & 12.8 & 100.0  \\
\hline
$\rm BR(\widetilde{\chi}_{2}^{0} \to \widetilde{\chi}_{1}^{0} \rm{h})(\%)$ 
& 66.1 & 0.0 & 0.0 & 77.9 & 87.2 & 0.0  \\
\hline
\hline
$\rm \sigma(p p \to{\widetilde{\chi}_{2}^{0}}{\widetilde{\chi}_{1}^{\pm}})_{\rm 8 \, TeV} ~(pb) $ 
& 0.671 & 0.662 & 0.190 & 0.196 & 0.111 & 0.110  \\
\hline
\end{tabu} 
\caption{Masses of stop, sbottom, neutralinos and charginos and the relevant branching fractions for the 
six benchmark points. All the other parameters are set to their fixed values as described in the text. 
In the rows 10-13 the branching fractions to the charginos and neutralinos have been summed over. 
In the final row, we give the pair production cross-section of lighest chargino and the second lighest neutralino for the 
8 TeV LHC. All the masses are in GeV.
\label{tab-1}}
\end{center}
\end{table}
%
It is worth mentioning that the cross-section for stop or sbottom pair 
production decreases rapidly with their increasing masses. For example, 
the next to leading order cross-section for a 500 GeV stop or sbottom 
pair production stands at around 680 fb (calculated using PROSPINO 
\cite{Beenakker:1996ed} with the default choices of the scale and parton 
distribution functions) and goes down to as low as 10 fb for masses of 1 
TeV even with the proton proton centre of mass energy of 14 TeV. Hence, 
to maximize the possibility to see a signal it is important to have most 
of the branching ratios contribute to the signal. As we mentioned in the 
previous section, this was precisely our motivation to consider a final 
state which is common to many of the decay chains of top and bottom 
squarks.

\section{Details of collider simulation and results}
\label{collider}

In this section we describe the details of our simulation procedure as 
well as the kinematic selection cuts for our signal and the 
backgrounds. But before we do that we would like to remind the readers 
about the final state we are interested in and also discuss the 
potential backgrounds for our signal. As we have already mentioned 
earlier, we are interested in a general search strategy for the third 
generation squarks and hence we consider the final state consisting of 
at least a top quark, at least two additional leptons and some missing 
transverse momentum. There are several Standard Model processes 
which can give rise to such final state and thus contribute to the 
background for our signal. The most important of these backgrounds are 
$\rm t \bar{t} + n \,\, jets$, $\rm t \bar{t} Z $, $\rm t \bar{t} W $ 
and $\rm t b W$. There are also other backgrounds like $\rm t \bar{t} t 
\bar{t}$ and $\rm t \bar{t} W W$ but they are expected to be 
insignificant because of their very low cross-section. In order to check 
that this is indeed true we also include them in our background 
estimation. We also check that the $\rm t W$ and $\rm t Z$ 
events, in spite of their comparatively larger cross-sections 
\cite{Campbell:2005bb,Campbell:2013yla}, do not contribute to the 
background. Note that, some of the backgrounds mentioned above, at the 
parton level, do not seem to contribute to the final state of our 
interest. For example, in $\rm t \bar{t} + n \,\, jets$ events once one 
top is tagged hadronically, there is no possibility of two additional 
leptons in the final state. However, once parton showering and hadronization 
is included, an additional lepton can be produced, for example, from a 
semileptonic B meson decay and thus this process can contribute to our 
final state.

We are now in a position to discuss the details of our event selection 
procedure. The reconstruction of final state objects like jets, leptons 
etc. along with the other selection criteria are described below.

\begin{figure}[ht!]
\begin{center}
\begin{tabular}{cc}
\includegraphics[width=240pt,height=220pt]{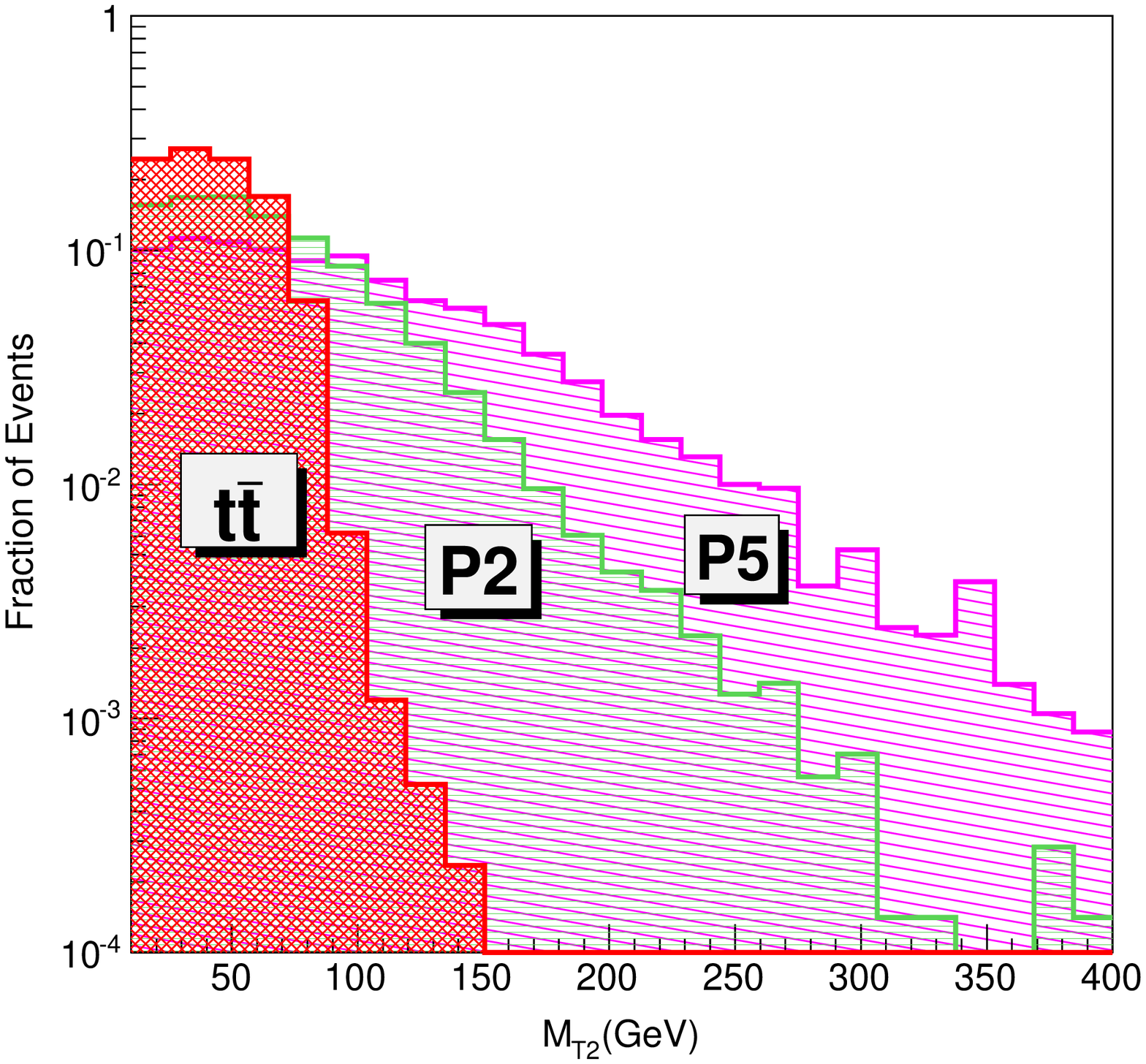}
\includegraphics[width=240pt,height=220pt]{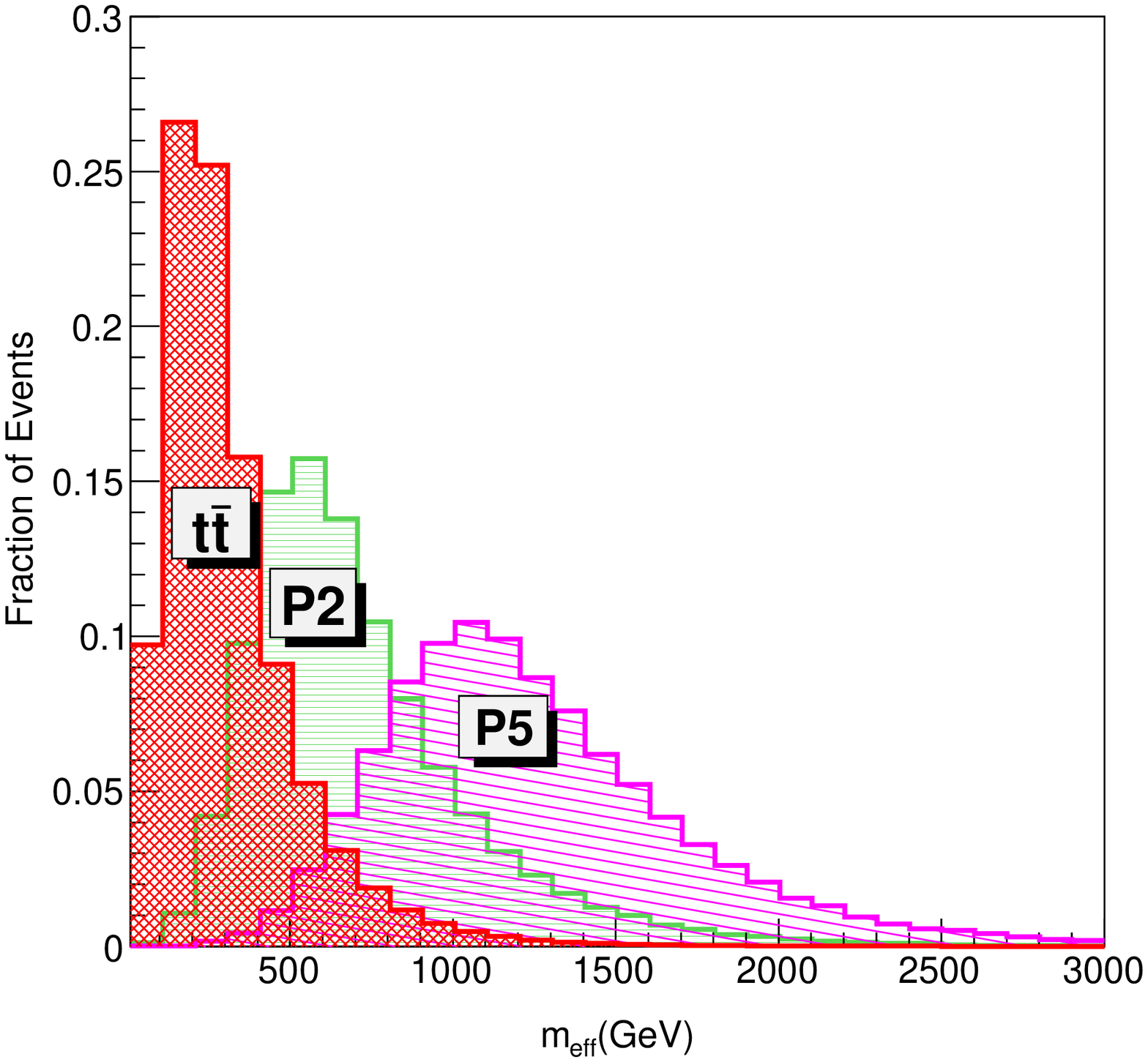}
\end{tabular}
\caption[]{The $\rm M_{T2}$ (left panel) and $\rm m_{eff}$ 
(right panel) distribution for PYTHIA generated $\rm t \bar{t}$ events
(marked as $\rm t {\bar t}$) and two 
benchmark points (denoted by {\bf P2} and {\bf P5}) 
with more than 50 $\rm fb^{-1}$ event sample.
\label{mt2-ht}}
\end{center}
\end{figure}

\begin{figure}[t!]
\begin{center}
\begin{tabular}{cc}
\includegraphics[scale=0.42]{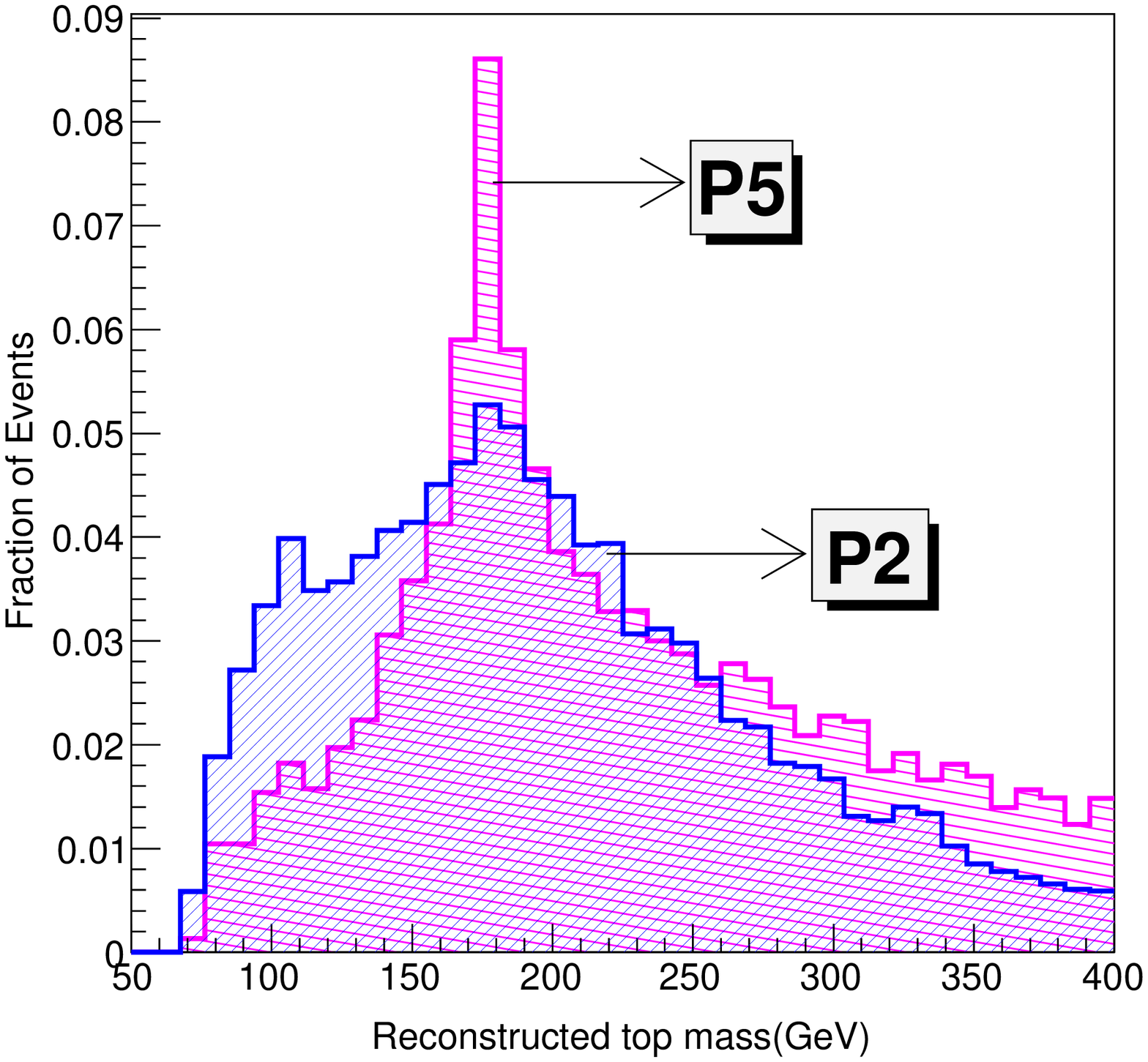}
\includegraphics[scale=0.42]{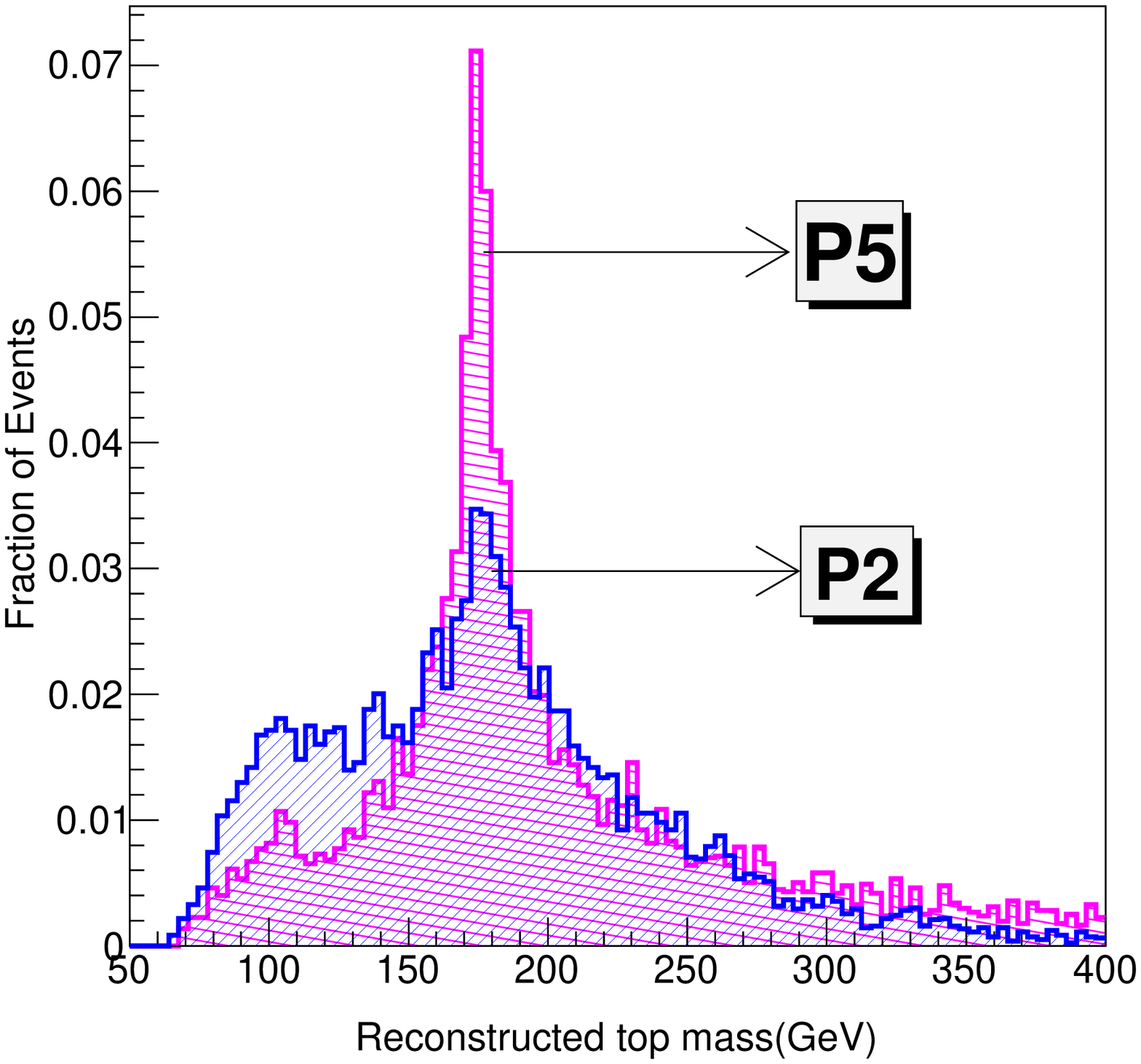}
\end{tabular}
\caption[]{The invariant mass distribution of the reconstructed top quark 
using the jet substructure algorithm without (left panel) and with 
(right panel) b-tag (assuming 60\% b-tagging efficiency) for two 
benchmark points with more than 50 $\rm fb^{-1}$ event sample.
\label{top-reco}}
\end{center}
\end{figure}

\begin{figure}[t]
\begin{center}
\vspace {-2.0cm}
\begin{tabular}{cc}
\includegraphics[scale=0.42]{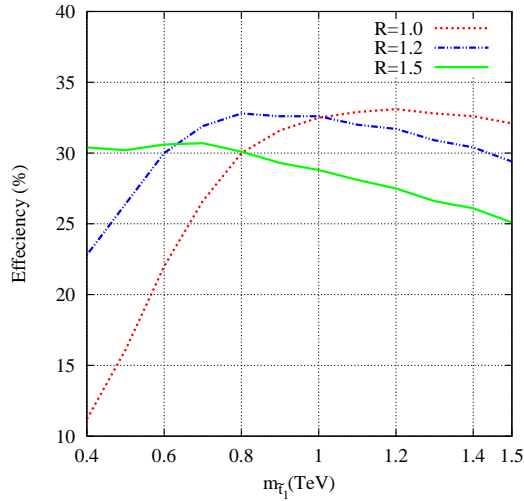}
\end{tabular}
\vspace {-2.0cm}
\caption[]{The top reconstruction efficiency as a function of stop mass 
for three choices of the parameter R.  
\label{stop}}
\end{center}
\end{figure}

\begin{itemize}

\item C1 : At first, we demand at least two isolated leptons (electron and muon) 
with the transverse momentum $\rm p_{T}^{\ell} \ge 25$ GeV and the 
pseudo-rapidity $|\eta| \le 3$. Isolation of leptons is ensured by 
demanding the total transverse energy $\rm p_T^{AC}$ to be less than 
20\% of $\rm p_T^{\ell}$. Here $\rm p_T^{AC}$ is defined as the scalar 
sum of transverse momenta of all the hard jets (j) which are close to 
the lepton satisfying $\rm \Delta R(\ell,j) \le 0.2$. The jets are 
formed using the anti-$\rm k_T$ algorithm with the value of R=0.5. Out 
of these jets only the hard jets which satisfy a transverse momentum cut 
$\rm p_{T}^j \ge 50$ GeV as well as the the pseudo-rapidity $|\eta| \le 
3$ are selected. The lepton identification at the LHC being extremely 
efficient and due to the clean environment, this can be used as an efficient 
trigger of our events. Our choice of the final state with two hard and 
isolated leptons also kills a large part of the SM background.

\item C2 : As a second step, we consider the variable $\rm 
M_{T2}$ \cite{Lester:1999tx,Barr:2003rg} which is defined as
\begin{equation} 
\rm M_{T2}(\overrightarrow{\rm p}_{T}^{\ell1}, 
\overrightarrow{\rm p}_{T}^{\ell2},\PMETV) \; = \; \rm 
min_{_{_{_{\hspace{-1cm}\Large \PMETV = \PMETV^{\, 1} + \PMETV^{\, 
2}}}}} \bigg[max\{M_T(\overrightarrow{\rm 
p}_{T}^{\,\ell1},\PMETV^{\,1}), M_T(\overrightarrow{\rm 
p}_{T}^{\ell2},\PMETV^{\, 2})\}\bigg],
\label{mt2def} 
\end{equation} 
where 
$\ell1$ and $\ell2$ are the two hard leptons selected in the previous 
step, $\rm \PMETV$ is the total missing transverse momentum vector of 
the event and $\rm M_T(\overrightarrow{\rm v}_{1}, \overrightarrow{\rm 
v}_{2})$ is the transverse mass of the $(\overrightarrow{\rm v}_{1}, 
\overrightarrow{\rm v}_{2})$ system which is defined as $$\rm 
M_T(\overrightarrow{\rm v}_{1}, \overrightarrow{\rm v}_{2}) = \sqrt{2 
|\overrightarrow{\rm v}_{1}|\,|\overrightarrow{\rm v}_{2}| (1 - 
\cos\phi)},$$ $\phi$ being the (azimuthal) angle between 
$\overrightarrow{\rm v}_{1}$ and $\overrightarrow{\rm v}_{2}$. In the 
definition of Eq.\ref{mt2def}, $\rm \PMETV^{\, 1}$ and $\rm \PMETV^{\, 
2}$ are a hypothetical split of the total observed missing transverse 
momentum into two parts. Here we have assumed the masses of the 
unobserved particles to vanish \cite{Barr:2009wu}. As for $\rm t \bar{t}$ 
events with both the top quarks decaying semi-leptonically 
the $\rm M_{T2}$ distribution is bounded above by the W boson
 mass, a hard cut on $\rm M_{T2}$ can help reduce 
the $\rm t \bar{t}$ background by a significant amount. 
This can be seen in the 
left panel of Fig.\ref{mt2-ht} where the $\rm M_{T2}$ 
distribution for $\rm t \bar{t}$ events as well as two signal points are 
shown. In our analysis we impose $\rm M_{T2} > 125$ GeV on 
the events.

\item C3 : We now define an effective mass of the system $\rm m_{eff} = 
\Sigma p_{T}^{j} + \Sigma p_{T}^{\ell}$, where the first sum runs over 
all the hard jets and the second sum is over all the hard and isolated 
leptons present in an event. As $\rm m_{eff}$ is strongly correlated 
with $2 \rm m_{\tilde{t}_1}$ or $2 \rm m_{\tilde{b}_1}$, the signal is 
expected to occupy the large $\rm m_{eff}$ region of the phase space 
while the backgrounds should have lower values of $\rm m_{eff}$. This 
can be seen in the right panel of Fig.\ref{mt2-ht} where the $\rm 
m_{eff}$ distribution for $\rm t \bar{t}$ and two signal points are 
shown. In our analysis, we find a lower cut of $\rm m_{eff} > 800$ GeV 
very useful to keep the backgrounds under control while keeping a 
significant number of signal events.

\item C4: As our signal final state consists of a number of stable 
neutralinos and neutrinos, a moderately hard missing transverse momentum 
cut $\rm \PMET > 150$ GeV is used to further reduce the backgrounds. 
This step helps to decrease, in particular, the $\rm t \bar{t} Z$ 
background to a good extent.

\item C5 : Finally, we demand at least one top quark in the sample by 
reconstructing its invariant mass using the jet substructure technique. 
For this purpose we use the Johns Hopkins top tagger\cite{Kaplan:2008ie},
using Cambridge Aachen (C/A) algorithm \cite{Dokshitzer:1997in}
with the choice of the parameters $\rm  R =1.5$, $\rm 
\delta_p=0.10$ and $\rm \delta_r=0.19$. While reconstructing W mass in 
the intermediate step of the algorithm we consider the mass window 
(60-100) GeV and demand that the W helicity angle $\theta_{h}$ satisfies 
cos$\theta_{h} < 0.7$. We demand the final reconstructed top mass to 
fall in the window (135-215) GeV. We do not impose a b-tag criteria on 
our reconstructed top quark.  The left panel of Fig.\ref{top-reco} shows 
the reconstructed top mass for two of our benchmark points. As the 
produced top quark is more energetic for heavier stop/sbottom quarks, 
the jet substructure technique for the top mass reconstruction works 
better for the benchmark-4. Although we have not used a b-tag criteria 
in our analysis, in the right panel of Fig.\ref{top-reco} we show the 
improvement of the mass reconstruction if a b-tag is demanded.
In Fig.\ref{stop} we also present the top reconstruction efficiency 
in the process 
$ \rm p \, p \to \tilde{t}_1 \, 
\tilde{t}_1^*, \, \, \tilde{t}_1 \to t \, \tilde{\chi}^{0}_2$ as a function 
of stop mass for three different values of R assuming the 
$\tilde{\chi}^{0}_2$ mass to be 150 GeV.
As expected, for R=1.0 and R=1.2, the efficiency initially increases with the 
increase in the stop mass, reaches a maximum and then starts decreasing for 
even larger values of the stop mass.
We demand that our choice of R should be optimal for a large
region of parameter space. From this figure it is clear that 
the choice of R=1.5 is adequate for our choice of stop mass range. 
The deviation in the efficiency for other two choices of the
parameter R is not more than few percent for our benchmark points.

\end{itemize}

We use Pythia6.4.24\cite{Sjostrand:2006za} for generating the signal 
events. For the backgrounds, we use Madgraph5 \cite{Alwall:2011uj} to 
generate parton level events and subsequently use the Madgraph-Pythia6 
interface (including matching of the matrix element hard partons and 
shower generated jets, following the MLM 
prescription\cite{Hoche:2006ph}) to perform the showering and implement 
our event selection cuts. We use the Fastjet3 
package\cite{Dokshitzer:1997in,Cacciari:2011ma} for the reconstruction 
of jets and the implementation of the jet substructure analysis for 
reconstructing the top quark.

\begin{table}[h!]
\small
\begin{center}
\tabulinesep=1.2mm
\begin{tabu}{|c|c|c|c|c|c|c|c|c|c|} 
\hline 
\multicolumn{3}{|}{} & 
\multicolumn{5}{|c|}{No. of events after the cut} & \\
\hline
Signal & Production  & Simulated events             
& C1 & C2   & C3 & C4 &  C5  & Final Cross-section \\
       &  Cross-section (fb)  & (in units of $10^4$)&    
        &      &    &    &      &  (in units of $10^{-2}$ fb)   \\
\hline 
P1 &  1130  & 10   & 10573 & 821 & 339  & 267 & 55  & 62.2 \\ 
P2 &  1130  & 10   & 11091 & 657 & 248 & 205 & 55 &  62.2 \\
P3 &   135  & 5    &  8043 & 1132 & 712 & 645 & 153 & 41.3 \\
P4 &   135  & 5    &  7713 & 1207 & 749 & 663 & 153 & 41.3  \\
P5 &    27  & 5    &  8623 & 1720 & 1414 & 1322 & 295 & 15.9  \\
P6 &    27  & 5    &  8543 & 1679 & 1343 & 1281 & 322 & 17.4  \\
\hline 
\end{tabu}
\caption{Event summary for the signal after 
individual cuts as described in the text. In the last column we show 
the final cross-section after all the event selection cuts have been 
applied.
\label{tab2a}} 
\end{center}
\end{table} 
\begin{table}[h!]
\small
\begin{center}
\tabulinesep=1.2mm
\begin{tabu}{|c|c|c|c|c|c|c|c|c|c|} 
\hline 
\multicolumn{3}{|}{} & 
\multicolumn{5}{|c|}{No. of events after the cut} & \\
\hline
SM backgrounds & Production  & Simulated events             
& C1 & C2   & C3 & C4 &  C5  & Final Cross-section \\
        &  Cross-section (fb)  & (in units of $10^4$)&    
        &      &    &    &      &  (in units of $10^{-2}$ fb)   \\
\hline 
$\rm t \bar{t} + $ jets &
 918000 \cite{Kidonakis:2011tg} & 4320  & 1587596  & 601  & 39  & 29  & 4  &  8.5 \\
$\rm t b W$ &
61000 & 600  & 215807 & 80  & 4  & 2  & 1 & 1.0 \\
$\rm t \, \bar{t}Z$ &
\phantom{cccc}1121 \cite{Kardos:2011na} & 7 & 6255   & 253 & 52 & 20 
& 2 & 3.2  \\
$\rm t \, \bar{t}W$ &
\phantom{cccc} 769 \cite{Campbell:2012dh} & 5 & 4471   & 31  & 3  & 2  
& 1 & 1.5  \\
$\rm t \, \bar{t}W^{+}W^{-}$ &
10 & 1 & 1588   & 33  & 14 & 13 & 6 & 0.6 \\
$\rm t \, \bar{t} \, t \, \bar{t}$ & 
10 & 1 & 1781   & 31  & 14 & 10 & 4 & 0.4 \\
\hline 
Total & \multicolumn{7}{|c|}{} &  \\
Background & \multicolumn{7}{|c|}{} & 15.2\\
\hline 
\end{tabu}
\caption{Event summary for the backgrounds after 
individual cuts as described in the text. In the last column we show 
the final cross-section after all the event selection cuts have been 
applied. For the $\rm t \bar{t} + $ jets background we have generated 
a matched sample of $\rm t \bar{t} + $ 0 jet, $\rm t \bar{t} + $ 1 jet and 
$\rm t \bar{t} + $ 2 jets using Madgraph.
\label{tab2b}} 
\end{center}
\end{table} 

We present our results in Table~\ref{tab2a} and \ref{tab2b} 
where the number of signal and background 
events surviving after each selection cut are shown respectively.
In Table~\ref{tab2a} and \ref{tab2b} the first three 
columns show the processes studied, the raw production cross-section and 
the number of events generated for each process respectively. For the 
signal points the raw cross-section corresponds to the 
next-to-leading-order value calculated using 
Prospino\cite{Beenakker:1996ed} with default choices for the scale and 
the parton density function. For the background processes we use either 
the NLO cross-sections if they available in the literature or the 
cross-section as given by Madgraph5. 
For both the signal as well 
as the backgrounds the total number of events simulated is of the same 
order or more than the number expected in the 14 TeV LHC with $\rm 50 
fb^{-1}$ luminosity. In the column 4-8 we show the number of events 
after each selection cut and the final column shows the cross-section 
after all the cuts have been imposed. A closer look on the numbers in 
Table~\ref{tab2a} shows how with the increasing masses of the stop and 
sbottom quarks from P1 to P4 (which makes the final state leptons, jets 
and missing energy more and more harder) the efficiencies of $\rm M_{T2}$ 
and $\rm m_{eff}$ increase for the signal events. These two steps also reduce 
the backgrounds to a manageable level. In the final step, the demand of 
a top quark in the sample brings down the background to a minuscule 
level keeping a handful of signal events.
\begin{table}[h!]
\small
\begin{center}
\tabulinesep=1.2mm
\begin{tabu}{|c|c|c|c|c|c|c|c|} 
\cline{3-8} 
\multicolumn{1}{c}{} & 
\multicolumn{1}{c|}{}& 
 \multicolumn{3}{|c|}{Signal($\rm N_S$) ( Background($\rm N_B$))} & 
 \multicolumn{3}{|c|}{{$ \rm{Significance}(\mathcal S) \; \rm{for} \;  \kappa = 10\% \, (30\%, 50\%)$}} \\ 
\hline
  & $ \rm m_{\tilde{t}_1}$(GeV) &  10 $\rm fb^{-1}$ & 50 $\rm fb^{-1}$ & 100 $\rm fb^{-1}$ & 10 $\rm fb^{-1}$ 
& 50 $\rm fb^{-1}$ & 100 $\rm fb^{-1}$ \\ 
\cline{1-8}
P1 & 501.6 & 6.2(1.6) & 31.1(8) & 62.2(16) & 4.9(4.6, 4.1) & 10.8(8.4, 6.3)  & 14.4(9.9, 6.9)  \\ 
P2 & 501.6 & 6.2(1.6) & 31.1(8) & 62.2(16) & 4.9(4.6, 4.1) & 10.8(8.4, 6.3)  & 14.4(9.9, 6.9)  \\ 
P3 & 714.2 & 4.1(1.6) & 20.7(8) & 41.3(16) & 3.2(3.0, 2.7) &  7.0(5.6, 4.2)  &  9.6(6.6, 4.6)  \\ 
P4 & 714.2 & 4.1(1.6) & 20.7(8) & 41.3(16) & 3.2(3.0, 2.7) &  7.0(5.6, 4.2)  &  9.6(6.6, 4.6)  \\ 
P5 & 918.1 & 1.6(1.6) &  7.9(8) & 15.9(16) & 1.3(1.2, 1.1) &  2.7(2.1, 1.6)  &  3.7(2.5, 1.8)  \\ 
P6 & 918.1 & 1.7(1.6) &  8.7(8) & 17.4(16) & 1.3(1.2, 1.1) &  2.9(2.3, 1.8)  &  4.0(2.8, 1.9)  \\ 
\hline 
\end{tabu}
\caption{The summary of our signal and backgrounds. Columns 3-5 show the number of signal 
(total background) events for three values of the integrated luminosity: 10 $\rm fb^{-1}$, 50 $\rm fb^{-1}$ and 
100 $\rm fb^{-1}$. The columns 6-8 show the statistical significance of our signal for the above three 
integrated luminosities. For each value of the integrated luminosity the significance is shown for three choices 
of the amount of possible systematic uncertainties, $\kappa=$ 10\%, 30\%  and 50\%. }
\label{tab3} 
 \end{center}
\end{table} 

In Table~\ref{tab3} we give a summary of our signal and backgrounds events 
and also estimate the signal significance ($\mathcal S$). While calculating the 
significance, in order to be conservative, we have also included a systematic uncertainty 
on the total background estimate. We thus define the significance $\mathcal S$ to be, 
\begin{equation}
\mathcal S = \frac{\rm N_S}{\sqrt{\rm N_B + (\kappa N_B)^2}} \, \, ,
\end{equation}
where $\rm N_S$  and $\rm N_B$ are the number of signal and background events respectively and 
$\kappa$ is the measure of the systematic uncertainty. We will show our results for three 
choices of $\kappa$, 10\%, 30\% and 50\%. 

Note that in principle, a detailed detector simulation has to be performed to check the robustness of 
our analysis. However, a true detector simulation is completely outside the scope of this work.  
In order to roughly estimate the detector effects on the efficiencies of the individual cuts in Table~\ref{tab2a} 
we have made our Madgraph/Pythia generated events for some of the signal points pass through the 
public detector simulation package Delphes\cite{Ovyn:2009tx} using the default ATLAS and CMS detector simulation cards 
provided by them. For the benchmark point P5, for example, we observe that the percentage of events 
which survived after the cut-2, 3, 4 and 5 are 17\%(16\%),  93\%(94\%), 91.8\%(88.5\%) and 19\%(21.8\%) 
respectively using the ATLAS(CMS) detector card as compared to 20\%,  82.5\%, 93.5\% and 22.3\% calculated 
from Table~\ref{tab2a}. We get similar results also for the other benchmark points as well as the backgrounds. 
Hence, we expect that the detector effects will not change our results to a significant extent and the  
uncertainty in our results due to these effects are safely taken care of by including the systematic uncertainty 
as mentioned in the previous paragraph. 

Before we move to our next section, we would like to breifly discuss on 
relevance of soft radiation and pileups effects on the jet substructure 
algorithm. It is well known that at higher collision energy 
and high luminosity run of the LHC, in the hadronic final states 
it will be very challenging to isolate high $p_T$ events 
in the  presence of the large number of additional soft pp collisions, 
pileup (PU) that occur simultaneously with any hard interactions. 
These low $p_T$ phenomena will adversely affect the 
absolute energy measurements of jets. Hence, it is of particular interest 
to understand the sensitivity of large size jets in presence of pileup. 
Grooming techniques may serve to
mitigate pileup sensitivity by effectively reducing the jet area.
Jet grooming methods (e.g., filtering \cite{Butterworth:2008iy},
trimming\cite{Krohn:2009th}, pruning\cite{Ellis:2009su}) were designed 
exactly for this purpose to remove the soft
uncorrelated radiation from the fat jet while retaining the final state
radiation off the resonance. Some recent analysis based on 
the leading order parton-shower Monte Carlo have been seen to agree 
with the data quite well \cite{Chatrchyan:2013rla,Aad:2013gja} after these 
grooming algorithms were used. Although it is impossible for us to simulate 
events including pileup and provide a quantitative estimate of its effect, 
we believe that the pileup contamination will not change our results in a 
significant way, in view of the efficient grooming techniques which seem 
to work pretty well even on the real data.  
\section{Interpretation in simplified scenarios}
\label{simplified}

In this section we interpret our results in two simplified model 
scenarios. In the first model we consider direct stop pair production 
and their decay through the decay chain 
$$ \rm p \, p \to \tilde{t}_1 \, 
\tilde{t}_1^*, \, \, \tilde{t}_1 \to t \, \tilde{\chi}^{0}_2, \, \, 
\tilde{\chi}^{0}_2 \to \tilde{\chi}^{0}_1 \, Z \, , $$ 
while in the 
second case we study the direct sbottom pair production using the 
following decay chain 
$$ \rm p \, p \to \tilde{b}_1 \, \tilde{b}_1^*, \, 
\, \tilde{b}_1 \to t \, \tilde{\chi}^{-}_1, \, \, \tilde{\chi}^{-}_1 \to 
\tilde{\chi}^{0}_1 \, W^{-} \, . 
$$

\begin{figure}[h!]
\begin{center}
\begin{tabular}{cc}
\includegraphics[scale=0.4]{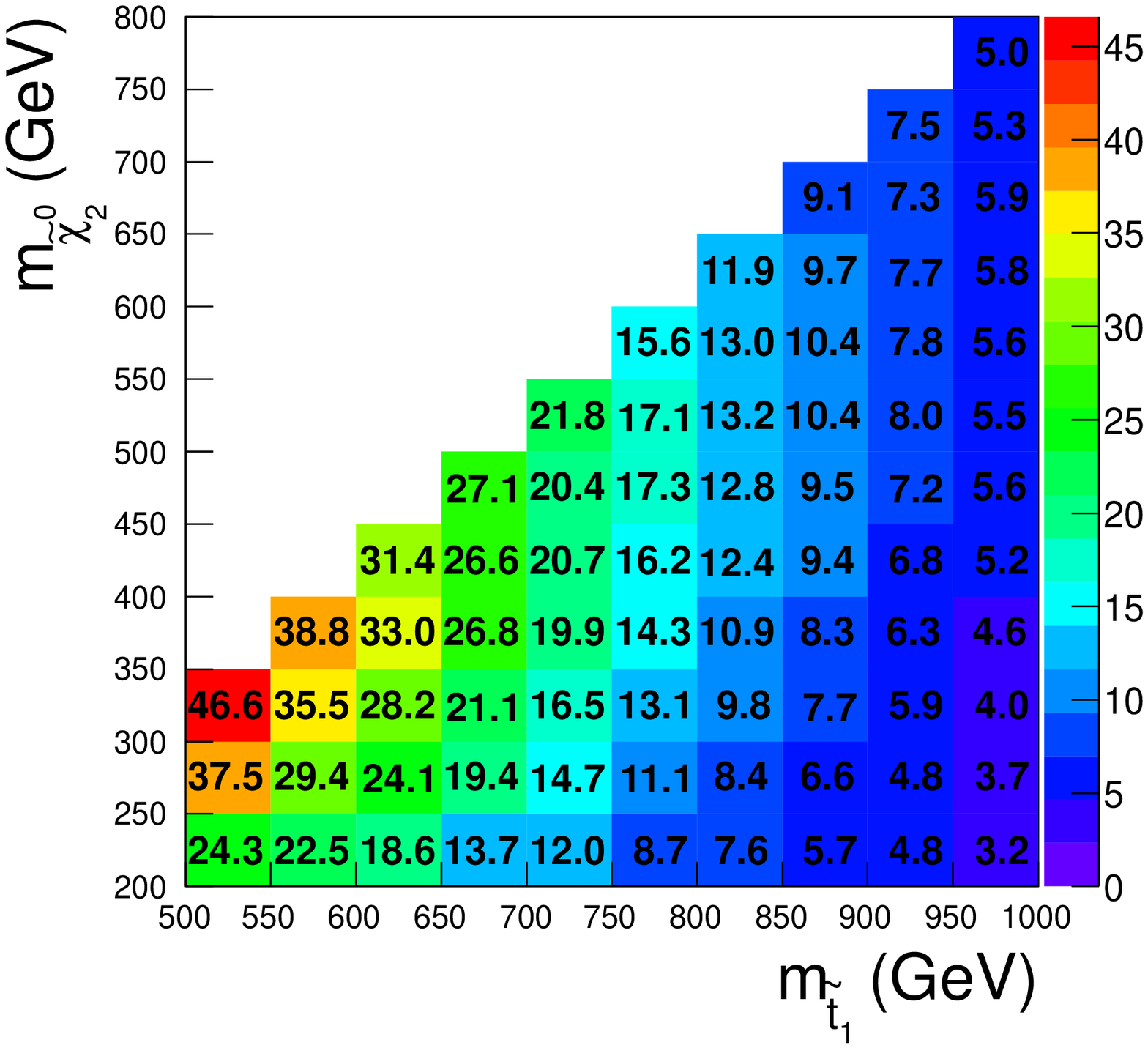}&
\includegraphics[scale=0.4]{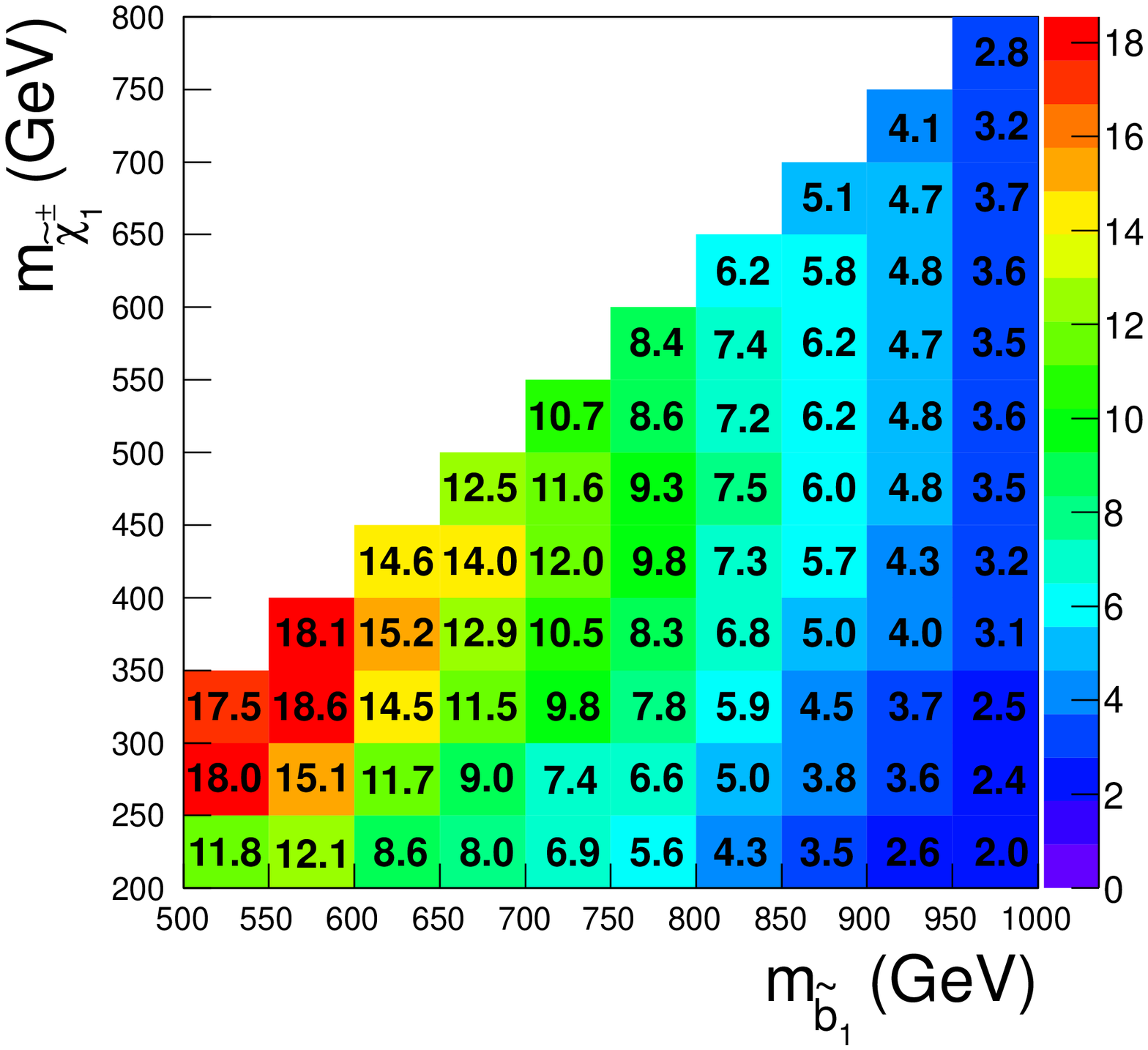}
\end{tabular}
\caption[]{The signal significance at 100 $\rm fb^{-1}$ in the 
$\rm \widetilde{t}_1$--$\rm \widetilde{\chi}_2^{0}$ plane (left panel) and 
$\rm \widetilde{b}_1$--$\rm \widetilde{\chi}_1^{\pm}$ plane (right panel) 
in the two simplified model scenarios as described in the text.
\label{simpl-1}}
\end{center}
\end{figure}

We assume all the branching ratios to be unity. In Fig.~\ref{simpl-1} we 
show the estimate of our signal significance calculated in the same way 
as in the previous section (with $\kappa$ = 10\%) corresponding to a 100 fb$^{-1}$ data set. 
The figure in the left panel shows the significance in the $\rm 
\widetilde{t}_1$--$\rm \widetilde{\chi}_2^{0}$ plane for our first 
simplified model while the figure in the right panel presents the 
significance in the $\rm \tilde{b}_1$-$\rm \tilde{\chi}^{\pm}_1$ plane 
in the second model. In both cases we assume the mass of the lightest 
neutralino to be 50 GeV. It can be observed that a top squark mass of 
about 1 TeV can be probed with 4-5$\sigma$ significance in the first case 
and sbottoms of mass about 900 GeV can be discovered with the same 
significance in the latter case.

\section{Summary and Conclusion}
\label{concl}

The non-observation of gluinos and squarks of first two generations and 
the discovery of a SM like higgs boson with mass around 125 GeV has 
forced us to consider the possibility of the existence of a light third 
generation of squarks namely, the top and the bottom quarks. As a light 
third generation of squarks is also motivated by naturalness arguments, 
searches of such light stop and sbottoms in all possible decay 
topologies are extremely important in the endeavour to look for signals 
of supersymmetry at the LHC. The production cross-section of a TeV scale 
stop or sbottom pairs being rather small, it is useful to consider an 
inclusive search strategy for stop and sbottom pair production in a 
common final state. In this work we attempt to perform such a study 
choosing a final state containing a top quark and two additional hard 
leptons along with substantial missing transverse momentum. We carry out 
a detailed simulation of the signal and all the possible backgrounds and 
observe that the combined use of di-leptonic $\rm M_{T2}$, the effective 
mass of an event $\rm m_{eff}$ and the jet substructure technique to tag 
a hadronically decaying top is extremely useful to achieve a good signal 
significance. We find that a third generation of squarks with masses 
about 900 GeV can be discovered at the 14 TeV LHC with a 100 fb$^{-1}$ 
data set which is achievable within the first few years of LHC14 run. We 
also interpret our results in the context of two simplified models of 
light stop and light sbottom squarks and conclude that with our strategy 
top squarks up to 1 TeV and bottom squarks up to about 900 GeV can be 
discovered with the same amount of data.

It is our hope that this work will motivate the experimental 
collaborations to perform further detailed analysis of the proposed 
channel on the real data taking our work as a starting guide.

\section{Acknowledgements}

DG acknowledges support from ERC Ideas Starting Grant n.279972 
``NPFlavour". DG thanks Satoshi Mishima for discussions and help in 
making some of the figures using ROOT. DG and DS also thank Monoranjan 
Guchait for encouragement and support and Sanmay Ganguly and Rajdeep M. 
Chatterjee of the CMS collaboration for discussion and help with some 
aspects of the software ROOT.

\input{3rdGen.bbl}

\end{document}

%% file: 3rdGen.bbl
\providecommand{\href}[2]{#2}\begingroup\raggedright\endgroup